\documentclass[12pt]{article}
\usepackage[dvips]{color}                 
\usepackage{cite,graphicx}                     
\usepackage{amssymb}
\usepackage{amsmath}
\usepackage{xspace}                       
\usepackage{epsfig}
\usepackage{setspace}

\newcommand{\mytitle}[1]{
                         \begin{center}
                           \LARGE{\textbf{#1}}
                         \end{center}}
\newcommand{\myauthor}[1]{\textbf{#1}}
\newcommand{\myaddress}[1]{\textit{#1}}
\newcommand{\mypreprint}[1]{\begin{flushright} #1 \end{flushright}}
\newcommand{\be}{\begin{equation}}
\newcommand{\ee}{\end{equation}}
\newcommand{\ba}{\begin{eqnarray}}
\newcommand{\ea}{\end{eqnarray}}

\begin{document}

\begin{titlepage}
\mypreprint{DESY 07-006 \\ IFUP-TH/2007-2}

\vspace*{0.5cm}
\mytitle{Decay constants  of charm and beauty pseudoscalar heavy-light mesons
on fine lattices}
  \vspace*{0.3cm}

\begin{center}
 \myauthor{A. Ali Khan$^a$},
 \myauthor{V. Braun$^a$}, 
 \myauthor{T. Burch$^a$}, 
 \myauthor{M. G\"ockeler$^a$},
 \myauthor{G. Lacagnina$^b$}, 
 \myauthor{A. Sch\"afer$^a$}
{\bf  and}
 \myauthor{G. Schierholz$^{c}$}

  \vspace*{0.5cm}
\myaddress{$^a$ 
 Institut f\"ur Theoretische Physik, 
 Universit\"at Regensburg, 93040 Regensburg, Germany} \\[2ex]
\myaddress{$^b$
  Dipartimento di Fisica, 
  Universit\`a di Pisa and INFN, \\ Pisa, Italy}\\[2ex]
\myaddress{$^c$
  John von Neumann-Institut f\"ur Computing NIC, Deutsches 
  Elektronen-Synchrotron DESY, \\ 15738 Zeuthen, Germany\\[2ex]
  Deutsches Elektronen-Synchrotron DESY, 22603 Hamburg, Germany}

  \vspace*{0.5cm}

QCDSF Collaboration

\end{center}

\vspace*{0.5cm}

\begin{abstract}
\noindent We compute decay constants of heavy-light 
mesons in quenched lattice QCD with a lattice spacing of $a \simeq 0.04$ fm
{\rm using}  non-perturbatively $O(a)$ improved Wilson fermions  {\rm and }
 $O(a)$ improved currents. {\rm We obtain}
$f_{D_s} = 220(6)(5)(11)$ MeV,
$f_D = 206(6)(3)(22)$ MeV, $f_{B_s} = 205(7)(26)(17)$ MeV and 
$f_B = 190(8)(23)(25)$ MeV, using the Sommer parameter $r_0 = 0.5$ fm 
to set the scale. The first error is statistical, the second systematic
and the third from assuming a $\pm 10\%$ uncertainty in the experimental 
value of $r_0$. A detailed discussion is given in the text.
We also present {\rm results} 
for the meson decay constants $f_K$ and
$f_\pi$ and the $\rho$ meson mass.

\vspace*{0.2cm}

\end{abstract}
\end{titlepage}

{\noindent \bf 1.} 
{\rm Weak decays of heavy-light mesons with $c$ and $b$ quarks are interesting for 
studies of CP violation and determination of the CKM mixing angles.
New experimental data on such decays are emerging 
(e.g.~\cite{Bellefb2006,CLEOfd2005,CLEOfds2006,Babarfds2006}) and 
their interpretation requires knowledge of hadronic matrix elements
governed by the strong interaction. Lattice QCD allows one to calculate
the strong matrix elements from first principles. However, 
if the heavy quark mass $m_Q$ is of the order of the inverse lattice spacing $a$,
considerable discretization effects proportional to
powers of $am_Q$ occur.}


One possibility for coping with this problem 
is to use an effective theory such as Heavy Quark Effective 
Theory (HQET)~\cite{eichten1989} or
Nonrelativistic QCD (NRQCD)~\cite{lepage1994}. These formalisms 
start from an infinitely heavy quark and 
consider corrections to this limit in the form of an
expansion in the inverse of  $m_Q$.  
However, to study the charm quark 
in HQET or NRQCD  requires a 
considerable number of correction terms, and one still has to worry about
the uncertainty from the truncation of the $1/m_Q$ expansion.
A formulation for relativistic quarks where masses can be of $O(1)$ in 
lattice units is the Fermilab approach~\cite{elkhadra1997} and 
modifications thereof, developed in~\cite{aoki2003,kayaba2006} 
and~\cite{lin2006}.
One expects that the dominating discretization effects are then proportional 
to powers of momenta of $O(\Lambda_{QCD})$.  
While within HQET non-perturbative renormalization is 
possible~\cite{dellamorte2006}, in
many of the calculations using effective theories the
renormalization constants are calculated only in perturbation 
theory (e.g. in Ref.~\cite{wingate2004}), leading to further
uncertainties.

Another possibility is to simulate on very fine lattices, and this
is the approach we have adopted in the present paper.
We have performed a quenched lattice study of heavy mesons 
with a lattice spacing $a$ of about 0.04 fm.
On such a fine lattice a relativistic treatment of the 
charm quark should be justified and we expect that 
discretization errors are small compared to
previous calculations on coarser lattices. 
We also make an attempt to study $B$ mesons in our relativistic framework.
{\rm
Even on our fine lattice we cannot simulate $B$ mesons directly, but
the required extrapolation becomes relatively short-range. We expect that the
resulting uncertainty is not much larger than the systematic error caused
by the use of an effective theory.
}
For example, recent unquenched 
calculations of $f_B$ and $f_{B_s}$~\cite{wingate2004,gray2005} employ NRQCD for the $b$ quark 
and quote a $\sim 10\%$ error based on
perturbation theory and other systematic effects.

In this article we present results 
for the leptonic decay constants of the $D_s$, $B_s$, $D$ and $B$ mesons.
We also evaluate light meson masses and decay constants to compare with
previous quenched calculations of the light spectrum on coarser lattices
and in order to be able to disentangle discretization and quenching effects.\\[2mm]


%
{\noindent \bf 2.} 
Our results are {\rm based on the analysis of}
114 quenched Wilson gauge configurations simulated at the coupling parameter $\beta = 6.6$
with a mixed heatbath and microcanonical overrelaxation algorithm
using the publicly available MILC code~\cite{milccode}. 
The lattice volume is $40^3 \times 80$, i.e.\ 
our lattice extends over $40$ points ($\sim 1.59$ fm) in space and 80 points in 
time. 
The lattice spacing is determined using the Sommer parameter $r_0 = 0.5$ fm.
This choice is motivated by a previous calculation~\cite{maynard2002}
which used $r_0$ to determine the lattice spacings and found that the 
results for $f_{D_s}$ from a 
quenched lattice and a lattice with $N_f = 2$ agreed 
($a \approx 0.1$ fm in these calculations).
From the interpolating formula given in~\cite{necco01},
one finds  for our lattice $a^{-1} = 4.97$ GeV.

For the quarks we use the $O(a)$ improved clover formulation~\cite{SW1985},
with the nonperturbative value of the clover coefficient  $c_{SW}= 1.467$ 
determined in Ref.~\cite{luscher1997a}. We work with seven  quark masses
corresponding to three ``light'' hopping parameters 
$\kappa = 0.13519, 0.13498, 0.13472$ and four ``heavy'' hopping parameters, 
$\kappa = 0.13000, 0.12900, 0.12100, 0.11500$. Statistical errors are
estimated by means of a bootstrap procedure using 500 bootstrap samples. 
For the central values
we take the median. The error bars are calculated including 34\% of the 
sample values below and above the median, respectively. Since the upper and
lower error bars are found to be quite symmetric for most of our data, we just 
quote the larger of the two. 
The autocorrelation times for the pseudoscalar meson propagator appear
to be small. In the worst case we studied, the autocorrelations decay after 
a distance of one configuration.

To extract the decay constants we follow the procedure described in 
Ref.~\cite{gockeler1998}.
For light and for heavy-light mesons we calculate the correlation functions 
\ba
C_{PA4}^{SL}(t) &=&  V \sum_{\vec x}
\langle A_4(\vec x, t) P^{S\dagger}(0) \rangle, \nonumber \\
C_{PP}^{Si}(t) &=&  V \sum_{\vec x} \langle P^i(\vec x, t) P^{S\dagger}(0) \rangle, 
\ea
where $A_4$ is the local axial vector current operator, 
$P$ the pseudoscalar density which can be local ($i=L$) or Jacobi 
smeared ($i=S$), and  $V$ is the spatial lattice volume.

\begin{table}[htb]
\begin{center}

\begin{tabular*}{0.90\textwidth}{@{\extracolsep{\fill}}|l|l|l|l|l|l|l|}
\hline
 \multicolumn{1}{|c|}{$\kappa_1 $}
& \multicolumn{1}{c|}{$\kappa_2 $}
& \multicolumn{1}{c|}{$am_{PS}$}  
& \multicolumn{1}{c|}{$am_V$}  
& \multicolumn{1}{c|}{$af^{(0)}$}  
& \multicolumn{1}{c|}{$af^{(1)}$}  
& \multicolumn{1}{c|}{$af$}  \\
\hline
 0.13519 &  0.13519 & $ 0.1059(13)$ & $ 0.1928(62)$  & $0.0376(11)$ & $0.0248(13)$  & $0.0312(09)$   \\
 0.13498 &  0.13519 & $ 0.1231(12)$ & $ 0.2005(48)$  & $0.0388(11)$ & $0.0251(12)$  & $0.0324(09)$   \\
 0.13498 &  0.13498 & $ 0.1388(10)$ & $ 0.2107(43)$  & $0.0403(10)$ & $0.0258(10)$  & $0.0337(08)$   \\ 
 0.13472 &  0.13519 & $ 0.1422(11)$ & $ 0.2065(39)$  & $0.0405(11)$ & $0.0261(11)$  & $0.0339(09)$   \\
 0.13472 &  0.13498 & $ 0.1560(10)$ & $ 0.2186(33)$  & $0.0418(10)$ & $0.0270(10)$  & $0.0351(08)$   \\
 0.13472 &  0.13472 & $ 0.1722(09)$ & $ 0.2292(27)$  & $0.0434(10)$ & $0.0283(09)$  & $0.0366(08)$   \\  
\hline
\end{tabular*} 
\end{center}
\caption{\small  Light meson masses  and decay constants in lattice units.}
\label{tab:lmasses}
\end{table}
Masses and amplitudes are determined from fits of the correlation functions 
with
\ba
C_{PP}^{Si}(t) & = & A_{PP}^{Si}
 \left(e^{-Et} + e^{-E(T-t)}\right),  \\
C_{PA4}^{SL}(t)  & = & A_{PA4}^{SL}  \left(e^{-Et} - e^{-E(T-t)}\right),
\ea
where $E$ is the ground state energy.
In Table~\ref{tab:lmasses} we give the raw data for the light 
pseudoscalar meson masses, determined from $C_{PP}^{SL}$, and for 
light vector meson masses from smeared-local 
correlation functions of the spatial components of the vector currents.

To determine the bare quark masses, we calculate
$\kappa_{crit}$, the $\kappa$ value corresponding to massless quarks, 
from a fit of the squared mass of a pseudoscalar meson (``pion'') 
consisting of quarks with mass parameters $\kappa_1$ and 
$\kappa_2$ as a function of the averaged $O(a)$ improved quark mass 
\be
\left(am_{PS}\right)^2 = a_1 \,a\tilde{m}_q, \label{eq:kappafit}
\ee
with
\[
\tilde{m}_q = (1 + b_m\,am_q)\,m_q\,,\; m_q =  \frac 1 2 (m_{q1}+m_{q2})
\]
and 
$am_{qi} = \frac 1 2 (\frac{1}{\kappa_i}-\frac{1}{\kappa_{crit}}),\,i=1,2$.
We use the non-perturbative value of $-0.6636$ for the improvement parameter
$b_m$ using an interpolating formula from Ref.~\cite{guagnelli2001}.
The fit includes all data with $\kappa_{1,2} \geq 0.13472$, where we find
the improved quark masses to just lie on a straight line.
We find $\kappa_{crit} = 0.135472(11)$. 
The hopping parameter corresponding to the average 
$u$ and $d$ quark mass, $\kappa_\ell$, is determined by
setting $m_{PS}$ on the left hand side of Eq.~(\ref{eq:kappafit}) 
equal to the
physical pion mass, $m_{PS} = 138$ MeV. We find $\kappa_\ell = 0.135456(10)$.

We parameterize the quark mass dependence of 
light meson  decay matrix elements 
with hopping  parameters $\kappa_1$ and $\kappa_2$
by fitting them to a function of the form
\be
c_0 + c_1 \, a\tilde{m}_q . \label{eq:chiralfit}
\ee
The light quark mass dependence of masses and decay matrix elements of heavy-light mesons
is parameterized using a linear fit as in Eq.~(\ref{eq:chiralfit}), with 
$\tilde{m}_q$ being the light quark mass instead of the average quark mass.

\begin{figure}[htb]
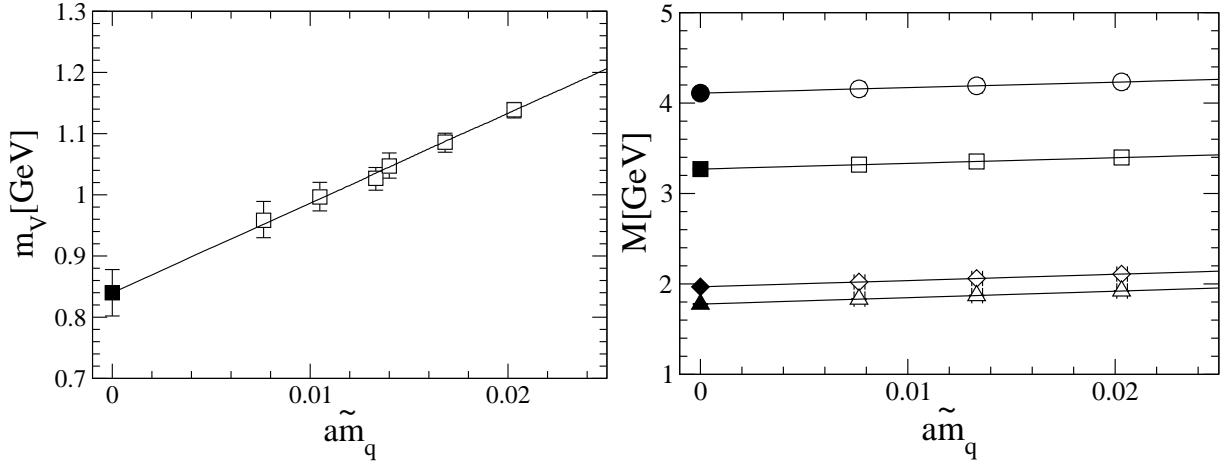

\begin{center}
\centerline{
\epsfig{file=rho.eps,width=8cm}
\epsfig{file=m_chiral.eps,width=8cm}
}
\end{center}
\vspace{-0.3cm}
\caption{\small Chiral extrapolation of  meson masses. On the left, light vector meson
masses, on the right, heavy-light pseudoscalar meson masses for the heavy hopping parameters
$\kappa = 0.115$ (circles),
 $\kappa = 0.121$ (squares),  $\kappa = 0.129$ (diamonds) and  
$\kappa = 0.130$ (triangles). 
Open symbols denote the simulation points, closed symbols the chiral extrapolation. }
\label{fig:hlmass}
\end{figure}

We also calculate the 
{\rm vector (``$\rho$ meson'')}
mass.
The fit and the chiral extrapolation assuming a quark mass dependence as in
Eq.~(\ref{eq:chiralfit}) are shown in Fig.~\ref{fig:hlmass}. At $\kappa_\ell$ we
find $846(37)$ MeV (the error is statistical), which is roughly a
$10\%$ ($2\sigma$) discrepancy with experiment.  
We compare our result to other recent
quenched calculations  in Table~\ref{tab:rhomasses}. Within errors our
result agrees with Ref.~\cite{pleiter_diss}, where a continuum
extrapolation from coarser lattices with $O(a)$ improved clover 
fermions is performed. We also list
studies employing chiral lattice fermions where smaller quark masses can be
reached while coarser lattices are 
used~\cite{gattr2004,galletly2006,bietenholz2006}. 
Ref.~\cite{galletly2006} quotes results from two lattice spacings.
In Table~\ref{tab:rhomasses} we present the results from their finer lattice.
To determine the strange quark mass parameter $\kappa_s$, we interpolate the vector meson
mass to the physical $\phi$ meson  mass, $M_\phi = 1.01946(19)$ GeV. 
We find $\kappa_s = 0.13502(6)$. This is our ``method I'' for determining the
$\kappa$ value corresponding to the strange quark mass. Using Eq.~(\ref{eq:kappafit})
and setting $m_K^2$ to the experimental value for $(m_{K_+}^2 + m_{K_0}^2)/2$ gives a value in
very close agreement: $\kappa_s = 0.134981(9)$.

The raw data for the heavy-light meson masses are given in 
Table~\ref{tab:hdecay}. To find the physical values of the heavy-light 
meson masses, we extrapolate for each heavy quark mass linearly in $m_q$,
see Eq.~(\ref{eq:chiralfit}). 
The fits are shown in Fig.~\ref{fig:hlmass}. 
The quark mass dependence is linear to very good accuracy. This is
in contrast to the findings of, e.g., Ref.~\cite{chiu2005}.
In the final step, 
the calculation of the decay constants, the physical values of the 
$c$ and $b$ quark masses will be reached by interpolating or extrapolating the
heavy-light meson mass to the $D$ or $B$ mass and the heavy-strange 
meson mass to the $D_s$ or $B_s$ mass.

In a quenched calculation, different methods to choose the input for
determining physical parameters may give different answers. In order to
investigate the influence of this arbitrariness we also use the 
heavy-light spectrum to determine $\kappa_s$ and call this
procedure ``method II''. 
We consider the splitting between mesons with a heavy quark and a 
strange quark (generically denoted by $M_s$) and a meson with a 
heavy quark and a quark with the $u,d$ quark mass (denoted 
by $M_\ell$). In our data, as well as in experiment, the $M_s-M_\ell$ 
mass difference is fairly independent of the heavy quark mass.
To fix  $\kappa_s$ in method II, we choose a heavy quark 
close to the charm mass from our simulation points, namely $\kappa= 0.129$, 
and set the splitting between the $M_s$ and the $M_\ell$ masses 
equal to the experimental value for the $D$ meson, $m_{D_s}-m_D = 98.85(30)$ MeV. 
The corresponding value for the strange hopping parameter is
$\kappa_s = 0.134929(15)$. 
%


%
%
%
\begin{table}[htb]
\begin{center}

\begin{tabular*}{0.90\textwidth}{@{\extracolsep{\fill}}|l|l|l|l|l|}
\hline
 \multicolumn{1}{|c|}{Ref.}
& \multicolumn{1}{c|}{$a^{-1}$[GeV]}
& \multicolumn{1}{c|}{quark action}
& \multicolumn{1}{c|}{gauge action}
& \multicolumn{1}{c|}{$m_\rho$[GeV]} \\ 
\hline
this work             &  $4.97$& clover            & Wilson           & $0.849(38)$   \\
\cite{pleiter_diss}   &  cont  & clover            & Wilson           & $0.797(13)$   \\
\cite{gattr2004}      & $1.33$ & chirally improved & L\"uscher-Weisz  & $0.791(42)$   \\  
\cite{gattr2004}    & $  1.29$ & fixed point       & fixed point      & $0.828(25)$   \\
\cite{galletly2006}   & $2.09$ &      overlap      & L\"uscher-Weisz  & $0.79(2)  $   \\ 
\cite{bietenholz2006} & $1.60$ & overlap hypercube & Wilson           & $1.017(40)$   \\
\hline
\end{tabular*} 
\end{center}
\caption{\small $\rho$ meson masses from quenched lattice calculations. 
The lattice scale has been determined using $r_0 = 0.5$ fm in all calculations
except in \cite{galletly2006} where $r_0 = 0.56$ fm is used. The quoted errors
are only statistical.}
\label{tab:rhomasses}
\end{table}
We calculate the pseudoscalar decay constants from the improved axial vector 
current $A_\mu^I$
\be
A_4^I = Z_A(1+ ab_A m_q) 
\left(A_4 + c_A a\partial_4 P\right), \label{eq:ren}
\ee
where $A_\mu(x) = \overline{q}_{1x} \gamma_\mu\gamma_5 q_{2x}$ and 
$P(x) =  \overline{q}_{1x}\gamma_5 q_{2x}$.
We take the nonperturbatively determined values for $Z_A$ from
 \cite{luscher1997b} and for $c_A$ from \cite{luscher1997a}. For our 
calculation, this gives
$Z_A = 0.8338$ and $c_A = -0.01967$. 
The coefficient $b_A$ is calculated from 1-loop perturbation 
theory~\cite{sint1996}. Using a boosted coupling 
$g_0^2 \rightarrow g_0^2/u_0^4$ with
$u_0 = \langle \frac 1 3 Tr U_P\rangle^{1/4}$, we find 
$b_A =  1.2143$ which is close to the result one finds using the tadpole-improved
scheme of~\cite{bhattacharya2001}. 
A non-perturbative determination of $b_A$
on coarser lattices ($\beta \leq 6.4$)~\cite{bhattacharya2006}  also
gives values in agreement with boosted perturbation theory within errors.

The meson matrix elements of the currents 
\ba
f^{(0)} & = & \frac{1}{M}\langle 0|A_4|M \rangle,\nonumber \\
f^{(1)} & = &  
\frac{1}{M}\langle 0|a\partial_4 P|M \rangle = 
-\frac{1}{M}\sinh(aM)\langle 0|P|M \rangle,\nonumber \\
f\hphantom{^{(0)}} & = & \frac{1}{M}\langle 0|A_4^I|M \rangle,
\ea
are related to the amplitudes by
\ba
f^{(0)} & = & -2\sqrt{\kappa_1\kappa_2}\frac{\sqrt{2}A_{PA4}^{SL}}
{\sqrt{MVA_{PP}^{SS}}} , \\
f^{(1)} & = & 2\sqrt{\kappa_1\kappa_2}\sinh(aM)\frac{\sqrt{2}A_{PP}^{SL}}
{\sqrt{MVA_{PP}^{SS}}} ,
\ea
where $M$ denotes the meson mass. 
The convention for the factors of $\sqrt 2$ corresponds
to the normalization where $f_\pi \simeq 130$ MeV. 
\begin{table}[htb]
\begin{center}

\begin{tabular*}{0.95\textwidth}{@{\extracolsep{\fill}}|l|l|l|l|l|l|}
\hline
 \multicolumn{1}{|c|}{$\kappa_1 $}
& \multicolumn{1}{c|}{$\kappa_2 $}
& \multicolumn{1}{c|}{$am_{PS}$}  
& \multicolumn{1}{c|}{$af^{0}$}  
& \multicolumn{1}{c|}{$af^{1}$}  
& \multicolumn{1}{c|}{$af$}  \\
\hline
0.11500 & 0.13519   & $0.8363(15)$ & $ 0.0371(11)$  & $ 0.0532(17)$ & $ 0.0423(13)$  \\
0.12100 & 0.13519   & $0.6676(13)$ & $ 0.0417(14)$  & $ 0.0496(17)$ & $ 0.0432(14)$  \\
0.12900 & 0.13519   & $0.4065(11)$ & $ 0.0475(13)$  & $ 0.0417(13)$ & $ 0.0435(12)$  \\
0.13000 & 0.13519   & $0.3685(12)$ & $ 0.0478(13)$  & $ 0.0399(13)$ & $ 0.0431(12)$  \\
0.11500 & 0.13498   & $0.8431(12)$ & $ 0.0383(12)$  & $ 0.0551(19)$ & $ 0.0437(13)$  \\
0.12100 & 0.13498   & $0.6747(11)$ & $ 0.0429(12)$  & $ 0.0517(16)$ & $ 0.0446(12)$  \\
0.12900 & 0.13498   & $0.4145(10)$ & $ 0.0488(15)$  & $ 0.0431(14)$ & $ 0.0448(14)$  \\
0.13000 & 0.13498   & $0.3765(09)$ & $ 0.0490(13)$  & $ 0.0412(12)$ & $ 0.0443(12)$  \\
0.11500 & 0.13472   & $0.8517(11)$ & $ 0.0402(12)$  & $ 0.0584(19)$ & $ 0.0460(13)$  \\
0.12100 & 0.13472   & $0.6836(10)$ & $ 0.0446(13)$  & $ 0.0541(16)$ & $ 0.0466(14)$  \\
0.12900 & 0.13472   & $0.4242(08)$ & $ 0.0508(13)$  & $ 0.0453(13)$ & $ 0.0469(12)$  \\
0.13000 & 0.13472   & $0.3866(08)$ & $ 0.0507(13)$  & $ 0.0430(12)$ & $ 0.0460(12)$  \\
\hline
\end{tabular*} 
\end{center}
\caption{\small Pseudoscalar heavy-light meson masses and decay constants at the simulation points.}
\label{tab:hdecay}
\end{table}
%
%

\vspace*{0.2cm}

\begin{figure}[htb]
\begin{center}
\epsfig{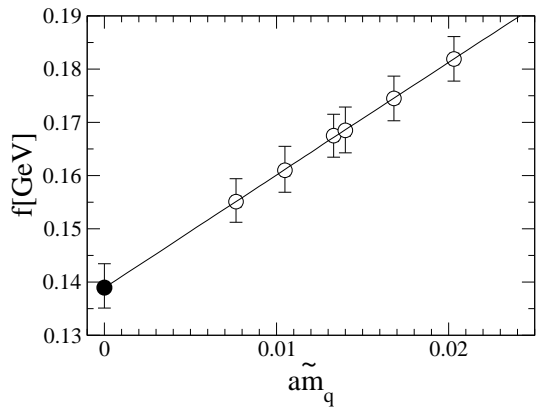}
\caption{\small Chiral fit of light meson decay constants. The 
chirally extrapolated value is denoted by the filled circle. }
\label{fig:l_decay}
\end{center}
\end{figure}

{\noindent \bf 3.} 
The fit of the light meson decay constants according to 
Eq.~(\ref{eq:chiralfit}) is shown in Fig.~\ref{fig:l_decay}. 
Mesons with degenerate and nondegenerate quark masses fall
on the same straight line.
For $f_\pi$, the value at the physical $u,d$ quark mass, and $f_K$, the value
extrapolated to the averaged strange and $u,d$ quark mass, we find
\be
f_\pi = 140(4) \mbox{ MeV}\,,\;\;
f_K   = 153(4) \mbox{ MeV}. 
\ee
This result for $f_\pi$ agrees well with the value of 137(2) MeV determined 
by~\cite{pleiter_diss}
using an extrapolation to the continuum from coarser lattices.
Both values are slightly larger than the experimental value of 
$f_{\pi^+} = 130.7(4)$ MeV. 
Our value for $f_K$ is $6\%$ or $2\sigma$ lower than the result
from~\cite{pleiter_diss} of 163(1) MeV.  The experimental value is
$f_{K^+} = 159.8(15)$ MeV.

The SU(3) flavor breaking ratio of the light decay constants in our 
calculation {\rm turns out to be}  relatively {\rm small.} 
We find 
\be
f_K/f_\pi -1 = 0.088(12)\,.
\ee
{\rm Our number is substantially lower}
than the experimental value of 0.222, but 
is consistent with a recent quenched calculation using overlap 
fermions~\cite{babich2006}, which finds $f_K/f_\pi -1 = 0.09(4)$ 
{\rm using the same scale setting} with
$r_0 = 0.5$ fm. It is also consistent 
with other quenched determinations (see~\cite{davies2004}).
%
%

\vspace*{0.2cm}

{\noindent \bf 4.}
{\rm Next we }
consider the heavy-light decay constants. 
To determine values at the physical quark masses, we extrapolate 
or interpolate the decay constants separately in the light and the 
heavy quark mass. For the fits in the light quark mass we use 
a function of the form (\ref{eq:chiralfit}) with $m_q$ being the
mass of the quark with the light $\kappa$ parameter of the
heavy-light meson instead of the average quark mass.
\begin{figure}[htb]
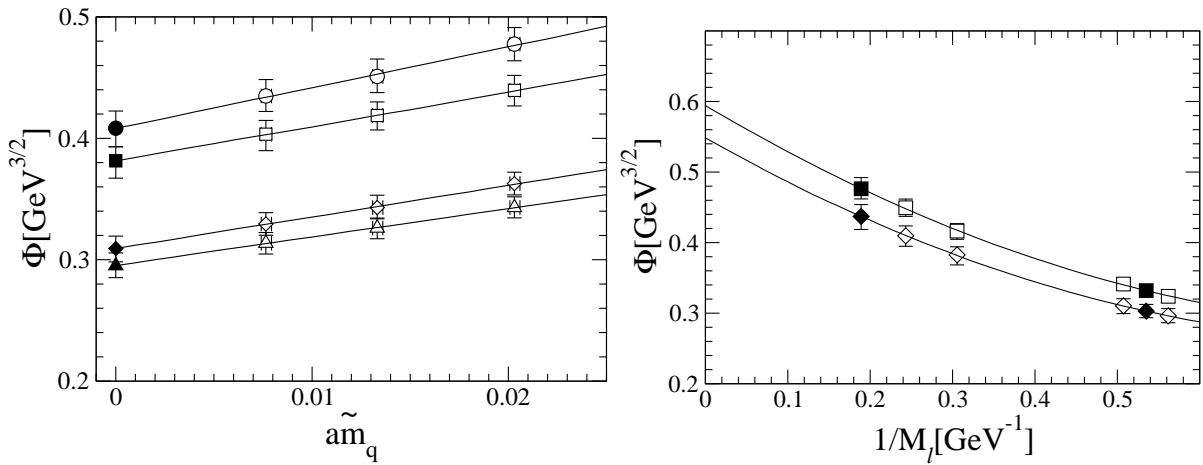

\begin{center}
\centerline{
\epsfig{file=Phi_chiral.eps,width=8cm}
\epsfig{file=Phi_vs_oneoverM_med.eps,width=7.75cm}
}
\end{center}
\vspace{-0.3cm}
\caption{\small On the left, chiral extrapolation of heavy-light decay matrix 
elements. Symbols have the same meaning as in the right part of 
Fig.~\ref{fig:hlmass}.
On the right, heavy quark mass dependence of 
heavy-light decay matrix elements. Squares 
denote strange, and diamonds denote physical light quarks. Closed symbols
denote heavy quark masses extrapolated to the $b$ or interpolated to the
$c$ quark mass.}
\label{fig:decay_vs_M}
\end{figure}
For the extrapolation to the $b$ quark mass and
also for an interpolation to the $c$ quark mass we use a formula motivated
by HQET (see e.g.~\cite{neubert1993}). 
{\rm In the heavy quark limit, matching of the decay matrix element in the effective 
theory to the matrix element in full QCD introduces logarithmic corrections in the heavy quark
mass which have to be resummed. In addition, power corrections in $1/m_Q$ have to be added.
Since the extrapolation to the $b$ mass in our case is rather 
short, the precise form of the extrapolation formula is not important. 
We use only the lowest order running for $\alpha_s$, 
and take the heavy-light meson mass $M_\ell$ (and not the quark mass $m_Q$) 
as an expansion (scale) parameter:
}
\be
\Phi \equiv \left(\frac{\alpha_s(M_B)}{\alpha_s(M_\ell)}\right)^{\gamma_0/(2b_0)}
\times f\sqrt{M_\ell} = c_0\left(1 + \frac{c_1}{M_\ell}+ \frac{c_2}{M_\ell^2}\right).
\ee
{\rm Here}
$\gamma_0 = -4$ is the leading {\rm order} 
anomalous dimension of the 
axial vector current, and $b_0 = 11$ is the leading 
coefficient of the {\rm QCD} $\beta$ function 
for zero dynamical flavors. 
The fits and the interpolated values are shown in Fig.~\ref{fig:decay_vs_M}.
The values of the fit parameters are $c_0 = 0.55(4)$ GeV$^{3/2}$,
$c_1 = -0.66(19)$ GeV and $c_2 = 0.38(21)$ GeV$^2$ 
if the light quark mass
is the $u,d$ quark mass, and $c_0 = 0.59(4)$ GeV$^{3/2}$,
$c_1 = -0.70(14)$ GeV and $c_2 = 0.39(15)$ GeV$^2$
for the $s$ quark.
\begin{table}[htb]
\begin{center}

\begin{tabular*}{0.80\textwidth}{@{\extracolsep{\fill}}|l|l|l|l|}
\hline
 \multicolumn{4}{|c|}{Decay constant ratios} \\
\hline
$f_{D_s}/f_{D}$ & $f_{B_s}/f_{B}  $ & $f_{D_s}/f_{B_s} $ & $f_{D}/f_{B}  $ \\
\hline
$1.068(18)(20)$ & $1.080(28)(31)$ &  $1.069(28)(160)$ & $1.082(42)(168)$ \\
\hline
\end{tabular*} 
\end{center}
\caption{\small  Ratios of heavy-light decay constants.  The first error is statistical, 
and the second systematic. The systematic errors are discussed in the text.}
\label{tab:results}
\end{table}
%
%

{\rm
Our final results for the ratios of heavy-light decay constants are presented in
Table~\ref{tab:results}, and the heavy-light decay constants are given along with 
a comparison in Table~\ref{tab:comp}.

Estimation of systematic errors is notoriously difficult. 
One source of uncertainty concerns setting the quark masses to their physical values.
For the strange quark this can be estimated by comparing the results from our methods I and II and
suggests an error of 4 MeV for $f_{D_s}$ and  $f_{B_s}$.
{}For the $u,d$ quarks a chiral extrapolation is required. The corresponding error is difficult 
to estimate. Our data are consistent with the simplest linear chiral extrapolation. 
Quenched chiral perturbation theory provides a more sophisticated formula.
However, it is not clear if it is applicable to our data.
The uncertainty in fixing the heavy quark mass can be estimated by comparing 
the difference between the mass fixed from quarkonium and 
from the heavy-light meson system. Since the $\eta_c$ meson mass 
using the charm quark hopping parameter determined from 
the $D_s$ meson agrees with the physical value, we assume that this uncertainty is rather
small in our calculation.
In addition, for the $B$ system there
is an uncertainty from the extrapolation in the heavy quark mass.
The difference between a quadratic fit to the matrix elements 
$f\sqrt{M}$ and a quadratic fit to $\Phi$ is very small and changes the
values for the decay constants by less than 1 MeV. If
only the three lighter heavy quark
masses are included in the extrapolation to the $b$ mass, $f_{B_s}$ ($f_B$) changes
by  $-3$ ($+1$) MeV.
}

Since we have results only from one lattice spacing, we cannot perform
a continuum extrapolation from our data alone and have to estimate the 
discretization effects as a systematic error.
Leading discretization effects are $O(a^2)$. 
{\rm A rough estimate of them can be obtained} 
by squaring the $O(a)$ corrections appearing
in the Symanzik improvement program. For the charm quark, the 
correction proportional to $c_A$ is small, around 2\%, while the term 
proportional to the quark mass and $b_A$ is around 10\% of the size 
of the matrix element itself. The square of the sum of these variations 
is around 1\%, which we take as our estimate for the discretization 
error of $f_D$ and $f_{D_s}$. A similar consideration for the 
$B$ and $B_s$ systems results in an
estimate of a discretization error of roughly $12\%$.  
For the error in the renormalization constants we use the estimate 
given for $Z_A$ in Ref.~\cite{luscher1997b} of 1\%.
\begin{figure}[thb]
\begin{center}
\epsfig{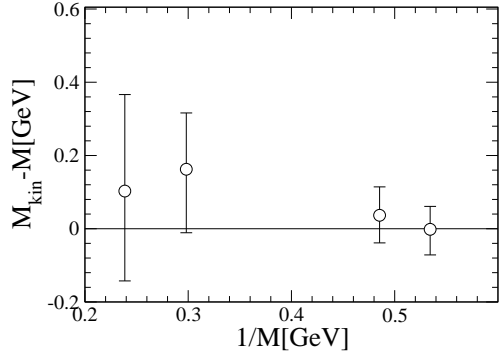}
\end{center}
\caption{\small Difference of the kinetic mass and the rest mass for heavy-light mesons 
with the light hopping parameter $\kappa = 0.13498$. The line at zero is plotted to guide the
eye. }
\label{fig:mkin}
\end{figure}
Since the heavy-light meson masses in lattice units in our simulation increase 
up to values of $\sim 0.8$ one {\rm might} be concerned about cutoff effects
in the dispersion relation for the heavy-light meson. 
We therefore compare the kinetic 
mass~\cite{elkhadra1997} $M_{\rm kin}$ calculated from 
\be
E^2 = M^2 + \frac{M}{M_{\rm kin}}\vec{p}^{\, 2} + O(p^4),
\ee
where $M$ is the rest energy of the meson. Results for the light quark mass close to
the strange quark mass are shown in Fig.~\ref{fig:mkin}.
We find that discretization errors in the dispersion relation are smaller
than the statistical errors.

The finite volume effects of the ratio $f_{B_s}/f_B$ have been 
investigated in the framework of heavy meson chiral perturbation 
theory~\cite{arndt2004}. For quenched lattices of spatial extent 
$1.6$ fm and pseudoscalar meson masses around $500$ MeV (which
corresponds to the smallest quark mass used in our simulation) they are quoted
to be around 1\% or smaller. It is plausible that the finite volume 
effects for $D$ and $D_s$ mesons are of similar size.

We estimate the
total systematic error due to discretization effects, errors in
$Z_A$, finite volume effects and ambiguities in fixing the physical quark masses by
collecting all contributions and adding them in quadrature. It is given as the
second error in Table~\ref{tab:comp}. 

The ratios are less sensitive to some of the systematic effects. The dominating ones
for $f_{D_s}/f_{B_s}$ and $f_{D}/f_{B}$ are the discretization effects. For $f_{D_s}/f_D$ and
$f_{B_s}/f_B$ we find a variation depending on how the strange quark mass is set,
while the estimated discretization effects are smaller than the statistical errors. 
The uncertainties from fixing the physical quark masses and the discretization errors
(added in quadrature) are given as the second error in Table~\ref{tab:results}.

We have used the  value of the 
Sommer parameter $r_0 = 0.5$ fm to set the scale in physical units. This choice allows 
for a direct comparison with previous lattice determinations (see below) but is not 
universally accepted. With a different value of the Sommer parameter our results have to
be modified accordingly. 
The variation of the
decay constants if $r_0$ is changed by $\pm 10\%$ is given as third error 
bar for our results in Table~\ref{tab:comp}~\footnote{We 
note, however, that a value of $r_0 = 0.45$ fm leads to seemingly unphysical results.
In particular the $SU(3)$ breaking in the meson masses and decay constants becomes very 
small. Also, different methods to set the strange quark mass 
produce more noticeable differences in the results. }.
%
\begin{table}[htb]
\begin{center}

\begin{tabular*}{0.90\textwidth}{@{\extracolsep{\fill}}|l|l|l|l|}
\hline
Ref. &  $N_f$,  HQ action, scale & $f_{D_s}$[MeV] & $f_{D}$[MeV]  \\
\hline
 \multicolumn{4}{|c|}{Lattice } \\
\hline
this work & 0,  clover, $r_0 = 0.5$ fm& $220(6)(5)(11)$ &  $206(6)(3)(22)$  \\
\protect\cite{giuseppe2000}  & 0,  clover, $r_0 = 0.5$ fm& 
$243(2)(^{03}_{24})$ & $222(3)(^{04}_{33})$ \\
\protect\cite{rolf2003}  & 0,  clover, $r_0 = 0.5$ fm & 252(9) &  \\
\protect\cite{juettner2005}  & 0,  clover, $r_0 = 0.5$ fm& 225(6) &  \\
\protect\cite{kayaba2006}  & 0, mod. Fermilab, $r_0 = 0.5$ fm& 237(5) &  \\
\protect\cite{chiu2005} &  0,  overlap, $f_\pi$ & 266(10)(18) & 235(8)(14) \\
\protect\cite{aubinfd2005} & $2+1$, Fermilab, $\Upsilon$ spectrum & 249(3)(16) & 201(3)(17) \\
\hline
 \multicolumn{4}{|c|}{Experiment } \\
\hline
\protect\cite{CLEOfds2006}   &  & $280(12)(6)$ &  \\
\protect\cite{Babarfds2006}  &  & $283(17)(16)$ &   \\
\protect\cite{CLEOfd2005}    &  & &  $223(17)(3)$ \\
\hline
\end{tabular*} 
\begin{tabular*}{0.90\textwidth}{@{\extracolsep{\fill}}|l|l|l|l|}
\hline
 Ref. & 
 $N_f$,  HQ action, scale & 
 $f_{B_s}$[MeV] &
 $f_{B}$[MeV] \\
\hline
 \multicolumn{4}{|c|}{Lattice } \\
\hline
this work & 0,  clover, $r_0 = 0.5$ fm & $205(7)(26)(17)$ &  $190(8)(23)(25)$  \\
\protect\cite{giuseppe2000}  & 0,  clover, $r_0 = 0.5$ fm& 
$240(4)(^{12}_{42})$ & $217(5)(^{13}_{40})$ \\
\protect\cite{dellamorte2005}  & 0,  clover+static, $r_0 = 0.5$ fm & 205(12) &  \\
\protect\cite{guazzini2006} &  0, clover+static, $r_0 = 0.5$ fm  & 191(6) &  \\
\protect\cite{wingate2004} &  $2+1$, NRQCD, $\Upsilon$ spectrum & 260(7)(28) &   \\
\protect\cite{gray2005} &  $2+1$, NRQCD, $\Upsilon$ spectrum &  &  216(9)(20) \\
\hline
 \multicolumn{4}{|c|}{Experiment } \\
\hline
\protect\cite{Bellefb2006} & experiment & &  $229(^{36}_{31})(^{34}_{37})$ \\
\protect\cite{utfit2006} & UTfit & 227(9) &  \\
 \hline
\end{tabular*} 
\end{center}
\caption{\small Heavy-light decay constants from lattice calculations and
experiment for the $D$  (upper table) and for the $B$ system (lower table). For the
lattice calculations, the number of flavors 
in the simulation ($N_f$), the heavy
quark (HQ) action, and the quantity used to set the scale are also indicated. 
The first error bar is the statistical, the second (where given) the systematic
error except for the uncertainty in $r_0$. For our work we quote a third error assuming
a $\pm 10\%$ uncertainty in the physical value of 
$r_0$. For the result from \cite{juettner2005} we quote the value from the
finest lattice instead of the continuum extrapolated result.}
\label{tab:comp}
\end{table}
%
%

\vspace*{0.2cm}

\begin{figure}[thb]
\begin{center}
\epsfig{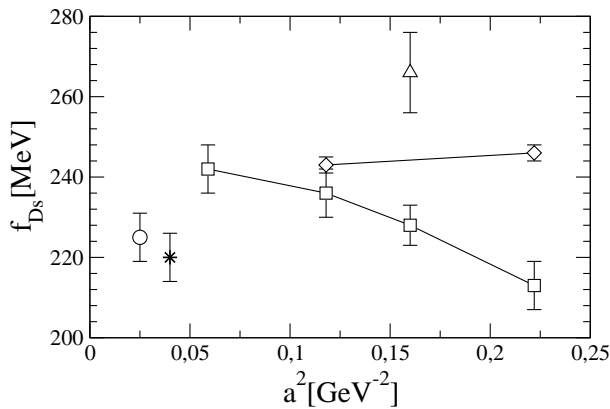}
\end{center}
\caption{\small Lattice spacing dependence of quenched $f_{D_s}$ from $O(a)$ 
improved clover 
quarks (this work, star), (UKQCD~\cite{giuseppe2000}, diamonds), 
(ALPHA~\cite{rolf2003}, squares), (J\"uttner~\cite{juettner2005}, circle),
and overlap quarks (Ref.~\cite{chiu2005}, triangle). 
The error bars show statistical and fitting uncertainties only. 
The scale is set using $r_0 = 0.5$ fm with the exception 
of~\cite{chiu2005} where  $f_\pi$ is employed for the conversion of the decay constant
to physical units.
If $r_0 = 0.5$ fm is used instead, their lattice spacing decreases by 
$12\%$,  which would increase their result for the decay constant 
even further. }
\label{fig:fDs}
\end{figure}

{\noindent \bf 5.}
{\rm Finally, we} compare our results to other lattice calculations of
decay constants. There exist recent quenched results for $f_{D_s}$ from nonperturbatively
$O(a)$ improved clover fermions~\cite{giuseppe2000,rolf2003,juettner2005}
for a range of lattice spacings ($0.03 \leq a \leq 0.1$ fm) as well as
for overlap quarks~\cite{chiu2005}. The comparison with the clover results
is particularly interesting because it sheds some light on the 
discretization effects and might indicate the possibility of
a joint continuum extrapolation.
We plot the clover data in Fig.~\ref{fig:fDs} as a 
function of the squared lattice spacing together with the overlap data. 
First we notice that on coarser lattices there is a
discrepancy between the clover data of Refs.~\cite{giuseppe2000} 
and~\cite{rolf2003}. The discrepancy 
corresponds roughly to the difference one obtains when $c_A$ values from different 
nonperturbative calculations for a meson mass $> 2.4$ GeV are used, as discussed in~\cite{giuseppe2000}.
Furthermore, {\rm the work}~\cite{rolf2003} 
uses a nonperturbatively determined value for $b_A$~\cite{bhattacharya2002}.
On the finer lattice of Ref.~\cite{giuseppe2000} ($\beta = 6.2$) the value used in 
Ref.~\cite{rolf2003}  is about $6\%$ larger than the perturbative 
number, which according  to our estimates would affect the decay constants by at most $2\%$. 
At $\beta = 6.0$ the difference
is even smaller.  
On the finest lattice used by~\cite{rolf2003}  ($\beta = 6.45$) 
the difference between the
perturbative and nonperturbative values of $b_A$ is $\sim 7\%$, which translates on a fine lattice into
only a very small difference in the decay constants. In a more recent 
calculation~\cite{bhattacharya2006} the nonperturbative value at that $\beta$ value has come into agreement
with perturbation theory, as mentioned in Section 2.

Our data is in good agreement with the value obtained by 
J\"uttner on his finest lattice~\cite{juettner2005}.
The overlap value from \cite{chiu2005} is on the other hand
substantially larger. Being determined on a relatively
coarse lattice it might be affected by discretization errors.
{\rm It is important that}
all data shown in Fig.~\ref{fig:fDs} come from lattices with {\rm similar} spatial extent 
between $1.5$ and $1.6$ fm. So, finite size effects {\rm can} be expected to be
{\rm roughly the same} in all calculations.

Our result for $f_{B_s}$ is consistent with the quenched 
calculations of~\cite{dellamorte2005,guazzini2006}, but considerably 
lower than the nonrelativistic (but unquenched) calculation of~\cite{wingate2004}.
The fit to the  standard model gives a value with a
relatively small error in between these two numbers.

{\rm For $f_D$ and $f_B$, the values obtained from lattice calculations 
are consistent with the experimental results.
Since the experimental errors are still large, 
this comparison is not conclusive, however.} 


\vspace*{0.2cm}

{\noindent \bf 6.} 
Let us summarize our main findings. 
We have calculated decay constants of heavy-light pseudoscalar mesons
on a very fine quenched lattice using clover fermions.
Our extrapolations to the $b$ quark mass appear reasonable. Nevertheless,
from a comparison of the results at the charm mass to 
data obtained on coarser lattices 
we obtain the impression that discretization errors with the relativistic
formalism adopted here are still significant for the $b$ sector, unless the 
inverse lattice spacing becomes larger than 
$\sim10$ GeV. 

Our results and those of Ref.~\cite{juettner2005} for $f_{D_s}$ 
are $10-15\%$ smaller than the central values quoted for other 
recent lattice calculations, and roughly $20\%$ smaller 
than recent experimental values. {\rm Eventually} one would like to determine the decay constants
to an accuracy of a few percent. {\rm Our} work and the result of Ref.~\cite{juettner2005}
indicate that 
discretization errors for the clover results on lattices with 
$a^{-1} \leq 2-3$ GeV are too large to reach this precision, 
and that even a continuum extrapolation from a set of coarser lattices has a large uncertainty  
for heavy quarks. {\rm On the other hand, we do not find any source of large systematic errors,
other than quenching, that could affect our calculation.} 
It seems, therefore, that 
the new lattice results on fine lattices (this work and~\cite{juettner2005}) 
indicate a relatively small  value for $f_{D_s}$ from lattice QCD. 
Quenching effects are notoriously difficult
to estimate.  However, since in previous calculations with $a^{-1} \approx 2$ 
GeV~\cite{maynard2002} it was found that the quenching error is insignificant
with our choice of lattice parameters, we expect that they will not be too
large. 

The systematic uncertainties on our results are larger for the $B$ system than
for the $D$ system and more difficult to estimate reliably. 
Our results are in agreement with several other recent lattice calculations, 
but smaller than the values from recent 
unquenched calculations using nonrelativistic methods.

We find a  rather small $SU(3)$ 
symmetry  breaking ratio of the heavy-light and 
light decay constants compared to experiment and also to several recent
unquenched lattice calculations. The difference between our numbers and the
unquenched results  may be partially 
due to the use  (see e.g.~\cite{aubinfd2005}) of a chiral 
extrapolation formula for the unquenched data which is 
inspired by chiral perturbation theory and predicts a particularly
strong decrease of
the decay constant at lighter quark mass values than are accessible in the
simulation. This is in contrast to the use of a simple linear extrapolation
in our calculation. 
%
\section*{Acknowledgements}
We would like to thank R. Sommer and A. J\"uttner for discussions.

The numerical calculations have been performed on the Hitachi SR8000
at LRZ (Munich). We are grateful to the LRZ for their support. 
We thank H. St\"uben for help with the inverter and its parallelization,
and D. Pleiter for help with the analysis program.

The work is supported in part by the DFG (Forschergruppe
Gitter-Hadronen-Ph\"anomenolo\-gie).
A.A. thanks the DFG and ``Berliner Programm zur F\"orderung der
Chancengleichheit f\"ur Frauen in Forschung und Lehre''
for financial support.



\end{document}